\documentstyle[12pt]{article}
\begin{document}
\date{}

{\Large
\begin {center}
Ferromaghetic detectors of axions in RF ($S\div X$) band
\end {center}
}
\vspace{0mm}
\begin {center}
P.V.\,Vorobyov, A.N.\,Kirpotin\\
Budker Institute of Nuclear Physics,\\
Novosibirsk,  Russia\\
Internet: vorobyov@inp.nsk.su\\
M.E.\,Rovkin\\
Tomsk Academy of Automated Control Systems and Radioelectronics,\\
Tomsk, Russia\\
A.P.\,Boldyrev\\
Sakhalin RadioPhysical Place, S-Sakhalinsk, Russia\\
\end {center}
\vspace{0mm}

\begin{quotation}{\em
We describe ferromagnetic detectors, for search of arion(axion),
where a high-sensitive two-channel SHF receiver is used.
Its sensitivity reaches to $10^{-20}\,Wt$,
with time of accumulation $1-10\,s$.

Fourier analysis of signal provides a survey in zone up to $\pm50\,KHz$
with spectral resolution $0.1 - 25\, Hz$.

There was applied a high sensitive SHF receiver based
on a special computer method of coherent accumulation of signals.
It is possible to use the receiver in other precise experiments:
measuring of electron/positron beams polarization in storage rings,
investigation of parity violation,
investigation of atmosphere with radars etc.
}
\end{quotation}
\vspace{4mm}

\section {Introduction.}

The (pseudo) Goldstone bosons arise naturally in many modern theories
such as supergravity, superstring theory
and variants of general relativity with torsion.
In the publications of last years the following (pseudo) Goldstone bosons
are discussing:
arion, axion, familon, majoron etc (see review \cite{b1}).

The experiments on the search of such interactions are interesting
from the point of view of fundamental physics (new long-range force)
and allow to obtain an information about \mbox{$10^6-10^{19}\,GeV$}-physics.
It is clear that this energy domain is inaccessible for accelerators.

By the other hand, there are well known indications
that a large part of the Universe mass exists in a form of dark matter.
The most attractive model of the dark matter
is non-relativistic gas of light elementary particles weakly interacting
with the "usual" matter \cite{b2} - \cite{b4}.

It means that a non-relativistic electron perceives the axionic
condensate as a space-inhomogeneous  magnetic  field  oscillating  in
time \cite{b3}.
The effective strength of this field is equal to:
\begin{equation}\label{Field}
{\bf B}_{eff}=2\kappa\sqrt{\rho_a}{\bf v}\sin
(m_at
+m_a{\bf v}{\bf x}+\theta).
\end{equation}
Here $\rho_a$ is the density of the condensate, $\kappa=\mu_a/\mu_B$,
$\mu_B$ is the Bohr magneton and $\theta$ is some phase.

The arion-electron interaction looks like
the very weak dipole magnetic interaction
and may be characterized by the "arionic magneton":
\begin{equation}
\mu_a=k\sqrt{{G_F}/8\pi}
\end{equation}
or $\mu_a=0.7\cdot 10^{-5}k\mu_B$.
Here $k^2=G_a/G_F$; $G_F$ is Fermi constant.

Let us suppose that ${\bf v}$ is equal to the "cosmological"
velocity of the Earth: $v\approx 10^{-3}$. Then the wavelength
corresponding to the space variations of the field ${\bf B}_{eff}$
can be estimated as
\begin{equation}\label{Length}
\lambda=0.1\left(\frac{1eV}{m_a}\right)\;(cm)
\end{equation}
If $m_a<1eV$ the length $\lambda \geq 0.1 cm$ and for samples of sizes
$\sim  1mm$  one  can  treat  the  field  ${\bf  B}_{eff}$  as   a
homogeneous
one:
\begin{equation}\label{Field-hom}
{\bf B}_{eff}={\bf b}\sin
(m_at+\theta),\;\;
{\bf b}=2\kappa\sqrt{\rho_a}{\bf v},
\end{equation}
where
$\kappa = \mu_a/\mu_b$.
Such an exotic quasimagnetic field with the amplitude about
$10^{-16} Gs$ can be picked up already in the present state of the art.
However, methods of its detection depend essentially on the axion mass
$m_a$ value being the frequency of the field's
${\bf B}_{eff}$ oscillations.

For massless arion \cite{b1}
the equation of motion for the arion field ($a$) and
electron spinor ($\psi$) are
\begin{equation}\label{Emae}
\begin{array}{l}
\Box a + \vec{\nabla} (\psi^{+} \vec{\sigma} \psi) = 0,\\
i\frac{\partial}{\partial t}\psi=\left[-\frac{\nabla^2}{2m}-q_{ea}
(\vec{\sigma}\cdot \vec{\nabla} a)\right]\psi
\end{array}
\end{equation}

The arion field may be described as the effective magnetic field
\begin{equation}
\vec{B_a}=-2\sqrt{\pi}\vec{\nabla}a
\end{equation}
The behavior of arion field in the wave zone has been considered
in \cite{b1}.
At large distances:
\begin{equation}
\vec{\nabla}a(\vec{r},t)=-\frac{\vec{r}}{r^2}
\left(\vec{n}\cdot\frac{q_{ea}}{8\pi}\int
\ddot{\vec{S}}(\vec{r}\prime,t\prime)
d^3\vec{r}\prime\right),
\end{equation}
where $\vec{n}=\vec{r}/r$,
$t\prime-|\vec{r}-\vec{r}\prime|$.
It is clear that in this case the arion wave the longitudinal polarization:
$\vec{\triangle}a\,\|\,\vec{n}$.
Such "ario-dynamics" is the base for constructing
arion/axion detectors.

At present time there are used
detectors based on coherent conversion
$axion(arion) \rightarrow photon$
in transverse magnetic field \cite{b5} - \cite{b7}.
For effective registration of axionic condensate, it is
advantageous to use a detector based on magnetized (anti)ferromagnetic.
In this case, axion-photon conversion happens
not through triangle diagram with charged fermion in loop,
but thorough process, analogous Compton-effect on electron bound in atom.
It increases $\alpha^{-2}$ times the probability of resonance
conversion axion-photon ($\alpha=1/137$) \cite{b8}.

So, as for search of cosmological axionic condensate,
as for building of a laboratory ferromagnetic detector based on
conversion \mbox{$magnon \rightarrow axion \rightarrow magnon$},
we need a high sensitive SHF receiver permitting
coherent accumulation of signal.

\section {SHF ferromagnetic detector of axions.}

If the quasimagnetic field has frequency  below  $10^6$  Hz  (what
corresponds
to $0<m_a<10^{-8}eV$) it is natural to use a detector with a ferromagnetic
rod as a sensitive body. Its magnetization can be read off by SQUID.
Detectors of such a kind have been used already in search of arion and
T-odd long-range forces \cite{b9} - \cite{b11}

In the range $10^8 Hz<m_a<10^{10}Hz$ ($m_a<10^{-4}eV$) one can
use the resonance axion-magnon conversion in magnetic ordered media
\cite{b8},\cite{b13}.
Let us consider a resonator with a working mode
$TE_{110}$ and with a small spherical ferrous- or antiferromagnetic
sample placed in its center. An external magnetic field is directed
in such a way that the averaged magnetization of the sample is
perpendicular to the magnetic component of the resonator's
eigenmode. The magnetic resonance frequency is fitted to be equal
to the eigenfrequency of the resonator. It provides a strong
coupling between the magnetic moment oscillations and the
electromagnetic ones. If $m_a$ coincides with this frequency
the spin waves will be exited resonantly by the axionic wind.
Electromagnetic oscillations coupled with such the waves can be
detected by a sensitive receiver.

Detailed discussion of the axion-magnon conversion and the corresponding
computations can be found in papers \cite{b8}, \cite{b12}.
Here we present the result only.
If $P$ is a limiting value of intensity which our receiver
can detect, $M$ is the sample magnetization, $Q_f$ is the quality of the
ferromagnetic resonance and $H_0$ is the external magnetic field, then
the smallest detectable quasimagnetic field is equal to:
\begin{equation}\label{Lim}
b\approx \left(\frac{P}{m_a^4}\right)^{1/2}\frac{H_0}{M}\frac{1}{kLQ_f}
\end{equation}
and for $P=10^{-15}erg/c$, $m_a=10^{10}Hz$, $H_0\sim M \approx 10^3 Gs$
and $(kL)^2=10^3$
we obtain $b\approx 10^{-15}Gs$. The use of the antiferromagnetic with a
large Dzyaloshinsky field (e.g. $Fe B O_3$) as a sensitive body can give
an additional factor $\sim 10^{-3}$ in the right hand side of (\ref{Lim})
(see \cite{b8}).

\subsection {A laboratory ferromagnetic detector}

Let us consider a laboratory ferromagnetic detector for search of
axions \cite{b8}.
Its main idea is following.
A powerful SHF generator excites the resonance precession of electrons'
spins in ferromagnetic, filling a waveguide-resonator.
Spin's coherent precession excites an axion wave which spreads along
waveguide axis and leaves it.
The axion wave penetrates freely through system of electromagnetic screens
gets into receiving waveguide
where it excites the resonance precession of spin in ferromagnetic.
The procession magnetic moment of ferromagnetic produces
an electromagnetic wave in the receiving waveguide-resonator;
High-sensitive coherent RF receiver registries this electromagnetic wave.

  We see, there happens the double conversion in detector:
\begin {displaymath}
{\it
\mbox {photon} \rightarrow
\mbox {magnon} \rightarrow
\mbox {axion}  \rightarrow
\mbox {magnon} \rightarrow
\mbox {photon}.
}
\end {displaymath}

The work of ferromagnetic detector
would be much more effective if we use a high sensitive receiver
providing a coherent accumulation of signal.

       The usage of the coherent detection scheme with quadrature component
registration leads to the square root of $N$ increase in the Signal-to-Noise
(SNR) ratio as compared to the non coherent scheme, where $N$ is the number
of accumulation cycles.

        The general idea of the detector is following:
The detector has a two-channel quadrature superheterodyne receiver
with a heterodyne, common for the both channels.
As a base generator, the heterodynes use a highly stable Rb frequency
generator.
The signal of ferromagnetic resonance (induced with axion field
in the receiving waveguide-resonator)
feeds to the first (high sensitive) receiver channel input,
is heterodyning, splits into its quadrature components and,
after the matched filters,
is registrating by couple of ADC's.\\
The "pilot-signal" of generator, through a "directed divider"
and attenuator, feeds to the second receiver channel input, and also
heterodynes, splits into its quadrature components and is digitizing
by second couple of ADC's.\\

The overall synchronization is provided by the
computer controlled generator of synchronization pulses.\\
As the accumulation is
completed, the data are read from the ADCs' buffers into the computer where
some sufficient processing takes place
(a statistical analysis of data in order to calculate zero levels
and other hardware parameters;
based on it, distinguishing of pure quadrature components
and compensation of their non-orthogonality;
reducing to the quadrature basis of the "pilot-signal" and
the further Fourier analysis).

{\em
Such detector, with operating frequency is $9\,GHz$, was constructed in BINP.
There was obtained following limit for the for electron-axion
coupling constant:
$q_{ae} < 10^{-3}\,GeV^{-1}$
or $B_{eff}/B_0 < 10^{-12}$ for distances less $5\,mm$.}

\subsection {HALOSCOPE - detector of cosmological axion condensate.}

Let us consider a detector of the cosmological axion condensate,
HALOSCOPE \cite{b3}.
It consists of SHF resonator, working in mode $TE_{011}$.
In the center of resonator, in the bunch of magnetic field,
there is placed a little spherical sample, made of (anti)ferromagnetic
of high ferromagnetic resonance quality.
The resonator is placed into constant external magnetic field so
that the sample's magnetization is perpendicular to
the magnetic component of resonator SHF field.

  We select such magnitude of external magnetic field that
the frequency  of  ferromagnetic  resonance  corresponds  the  own
frequency
of resonator at mode $TE_{011}$.

Let the detector move through the axion condensate so
that the vector of effective quasimagnetic field  $\vec{b}$
is perpendicular to the external field.
If the field oscillation frequency $\vec{b}$ coincides with the
the frequency of ferromagnetic resonance
then a precession of ferromagnetic magnetization vector
will take place.
There appears an oscillating component of the sample's magnetic moment;
it is well linked with the magnetic field $TE_{011}$ of the
resonator's oscillation.

  This linkage provides a transmission of power from quasimagnetic wave
to resonator's electromagnetic field (and back) which is possible
to registry by a high sensitive RF receiver.

   There are possible two ways to use the receiver:
\begin{itemize}
\item[1)]
The signal of ferromagnetic resonance
feeds to the high sensitive receiver input,
and is slitting into its quadrature components,
relatively basis formed by very stable heterodyne.
In this case, we do not need the low sensitive support channel
for "pilot-signal".
\item[2)]
A "pilot-signal" from very stable generator is using;
it feeds to a support channel input (low sensitive).
This way looks more advantageous because:
\item[-]
  We can use SHF-modules of standard SHF receivers like "πλ7-8",
\mbox {"πλ7-11"} and other ones
  which contains SHF-amplifier, heterodyne etc.
  A low stability of own first heterodyne does not matter
  because the coherentness is provided by high stable "pilot-signal"
  which is easy to obtain.
\item[-]
  It is not necessary for the "pilot-signal" to be, simultaneously,
  very stable and powerful enough.
\item[-]
  The "pilot-signal" may be used for adjusting of ferromagnetic
 resonance and for calibration of detector (among measurement cycles).
\end{itemize}

{\em
This detector was constructed in BINP
for operation in range $2-12\,GHz$.
We scanned range $8-10\,GHz$ with sensitivity $B_{eff} < 10^{-14}\,Gs$.
For this sensitivity level no axionic signal was found.
It is a preliminary result; we hope to perform a full scanning ($2-12\,GHz$)
with better sensitivity.

To increase sensitivity, we used in this work
the method of coherent accumulation and modulation of magnetized field.}

\section {High-sensitive SHF receiver}

We constructed a complex of tho-channel coherent receivers covering
scope from $100\,MHz$ to $12\,GHz$ \cite{b14}.\\
The receivers are made by classic heterodyne scheme with
double frequency transformation.
To provide the coherence, a computer processing was used.

For each channel, the signal is heterodyned and split into its inphase
and quadrature components, which are digitized by couple of ADC's and,
after accumulating cycle, are read by computer.\\
A special very effective procedure performs a statistical analysis of
accumulated sequences and calculate floating zeros
(of all tract of measure including ADCs) and parameters of non-orthogonality
of the quadrature splitter.
After this procedure, there becomes possible to obtain pure quadrature
components of signals and to compensate their non-orthogonality.

As the quadrature components are calculated, every couple of quadratures
from accumulating signal array is reduced to the quadrature basis
of corresponding couple of "pilot signal's" quadratures
or, the same,
random initial phases of "pilot signal" are subtracted from the phases
of accumulating signal.\\
So a coherent sequence of accumulating signal's quadratures is obtained;
and its Fourier analysis and other necessary processing are performing.

Using this method,
there was reached the sensitivity $-180\,dB$, while accumulating
sequence of length 4096.

Another complex was used in radar for investigation of atmosphere,
with work frequency of $150\, MHz$.
Its sensitivity is $-190\,dB$ \cite{b15}.

This work was supported from by "COSMION" foundation.

\end{document}